\documentstyle[12pt]{article}

\begin{document}

\author{Joaquim A. Dente \\
Laborat\'orio de Mecatr\'onica, Departamento de Engenharia \\
Electrot\'ecnica e Computadores, Instituto Superior T\'ecnico, \\
Av. Rovisco Pais, 1096 Lisboa Codex, Portugal \and Rui Vilela Mendes\thanks{%
phone: 351 1 7950790, fax: 351 1 7954288, e-mail: vilela@alf4.cii.fc.ul.pt} 
\\
Grupo de F\'\i sica-Matem\'atica, Complexo II, Univ. de Lisboa, \\
Av. Gama Pinto 2, 1699 Lisboa Codex, Portugal\\
and\\
Centre de Physique Th\'eorique, CNRS, Luminy, \\
Case 907, F 13288 Marseille Cedex 9, France}
\title{Characteristic functions and process identification by neural networks}
\date{}
\maketitle

\begin{abstract}
Principal component analysis (PCA) algorithms use neural networks to extract
the eigenvectors of the correlation matrix from the data. However, if the
process is non-Gaussian, PCA algorithms or their higher order
generalisations provide only incomplete or misleading information on the
statistical properties of the data. To handle such situations we propose
neural network algorithms, with an hybrid (supervised and unsupervised)
learning scheme, which constructs the characteristic function of the
probability distribution and the transition functions of the stochastic
process. Illustrative examples are presented, which include Cauchy and
L\'evy-type processes.
\end{abstract}

\section{Introduction}

Let $x_i$ denote the output of node $i$ in a neural network. Hebbian
learning (Hebb 1949) is a type of unsupervised learning where the neural
network connection strengths $W_{ij}$ are reinforced whenever the products $%
x_ix_j$ are large. If $Q$ is the correlation matrix 
\begin{equation}
\label{1.1}Q_{ij}=\left\langle x_ix_j\right\rangle 
\end{equation}
and the Hebbian learning law is local, all the lines of the connection
matrix $W_{ij}$ will converge to the eigenvector of $Q$ with the largest
eigenvalue. To obtain other eigenvector directions requires non-local laws
(Sanger 1989, Oja 1989, 1992, Dente and Vilela Mendes 1996). These principal
component analysis (PCA) algorithms find the characteristic directions of
the correlation matrix $Q$. If the data has zero mean ($\left\langle
x_i\right\rangle =0$) they are the orthogonal directions along which the
data has maximum variance. If the data is Gaussian in each channel, it is
distributed as a hyperellipsoid and the correlation matrix $Q$ already
contains all information about the statistical properties. This is because
higher order moments of the data may be obtained from the second order
moments. However, if the data is non-Gaussian, the PCA analysis is not
complete and higher order correlations are needed to characterise the
statistical properties. This led some authors (Softy and Kammen 1991, Taylor
and Coombes 1993) to propose networks with higher order neurons to obtain
the higher order statistical correlations of the data. An higher order
neuron is one that is capable of accepting, in each of its input lines, data
from two or more channels at once. There is then a set of adjustable
strengths $W_{ij}$ , $W_{ij_1j_2}$ , $\cdots $ , $W_{ij_1\cdots j_n}$ , $n$
being the order of the neuron. Networks with higher order neurons have
interesting applications, for example in fitting data to a high-dimensional
hypersurface. However there is a basic weakness in the characterisation of
the statistical properties of non-Gaussian data by higher order moments.
Existence of the moments of a distribution function depends on the behaviour
of this function at infinity and it frequently happens that a distribution
has moments up to a certain order, but no higher ones. A well-behaved
probability distribution might even have no moments of order higher than one
(the mean). In addition a sequence of moments does not necessarily determine
a probability distribution function uniquely (Lukacs 1970). Two different
distributions may have the same set of moments. Therefore, for non-Gaussian
data, the PCA algorithms or higher order generalisations may lead to
misleading results.

As an example consider the two-dimensional signal shown in Fig.1. Fig. 2
shows the evolution of the connection strengths $W_{11}$ and $W_{12}$ when
this signal is passed through a typical PCA algorithm. Large oscillations
appear and finally the algorithm overflows. Smaller learning rates do not
introduce qualitative modifications in this evolution. The values may at
times appear to stabilise, but large spikes do occur. The reason for this
behaviour is that the seemingly harmless data in Fig.1 is generated by a
linear combination of a Gaussian with the following distribution%
$$
p\left( x\right) =K\left( 2+x^2\right) ^{-\frac 12} 
$$
which has first moment, but no moments of higher order.

To be concerned with non-Gaussian processes is not a pure academic exercise
because, in many applications, adequate tools are needed to analyse such
processes. For example, processes without higher order moments, in
particular those associated with L\'evy statistics, are prominent in complex
processes such as relaxation in glassy materials, chaotic phase diffusion in
Josephson junctions and turbulent diffusion (Shlesinger et al 1993, Zumofen
and Klafter 1993, 1994).

Moments of an arbitrary probability distribution may not exist. However,
because every bounded and measurable function is integrable with respect to
any distribution, the existence of the characteristic function $f(\alpha )$
is always assured (Lukacs 1970). 
\begin{equation}
\label{1.2}f\left( \alpha \right) =\int e^{i\alpha x}dF\left( x\right)
=\left\langle e^{i\alpha x}\right\rangle 
\end{equation}
where $\alpha $ and $x$ are N-dimensional vectors, $x$ is the data vector
and $F(x)$ its distribution function.

The characteristic function is a compact and complete characterisation of
the probability distribution of the signal. If, in addition, one wishes to
describe the time correlations of the stochastic process $x(t)$, the
corresponding quantity is the characteristic functional (Hida 1980) 
\begin{equation}
\label{1.3}F\left( \xi \right) =\int e^{i\left( x,\xi \right) }d\mu \left(
x\right) 
\end{equation}
where $\xi \left( t\right) $ is a smooth function and the scalar product is 
\begin{equation}
\label{1.4}\left( x,\xi \right) =\int dtx\left( t\right) \xi \left( t\right) 
\end{equation}
where $\mu \left( x\right) $ is the probability measure over the sample
paths of the process.

In the following we develop an algorithm to compute the characteristic
function from the data, by a learning process. The main idea is that in the
end of the learning process we should have a neural network which is a
representation of the characteristic function. This network is then
available to provide all the required information on the probability
distribution of the data being analysed. To obtain full information on the
stochastic process, a similar algorithm might be used to construct the
characteristic functional. However this turns out to be computationally very
demanding. Instead we develop a network to learn the transition function and
from this the process may be characterised.

\section{Learning the characteristic function}

Suppose we want to learn the characteristic function $f(\alpha )$ (Eq. \ref
{1.2}) of a one-dimensional signal $x(t)$ in a domain $\alpha \in \left[
\alpha _0,\alpha _N\right] $ . The $\alpha $-domain is divided into N
intervals by a sequence of values $\alpha _0$ $\alpha _1$ $\alpha _2$ $%
\cdots $ $\alpha _N$ and a network is constructed with N+1 intermediate
layer nodes and an output node (Fig.3).

The learning parameters in the network are the connection strengths $W_{0i}$
and the node parameters $\theta _i$. The existence of the node parameter
means that the output of a node in the intermediate layer is $\theta _i\chi
_i\left( \alpha \right) $, $\chi _i$ being a non-linear function. The use of
both connection strengths and node parameters in neural networks makes them
equivalent to a wide range of other connectionist systems (Doyne Farmer
1990) and improves their performance in standard applications (Dente and
Vilela Mendes 1996). The learning laws for the network of Fig.3 are: 
\begin{equation}
\label{2.1}
\begin{array}{c}
\theta _i\left( t+1\right) =\theta _i\left( t\right) +\gamma \left( \cos
\alpha _ix\left( t\right) -\theta _i\left( t\right) \right) \\ 
W_{0i}\left( t+1\right) =W_{0i}\left( t\right) + \\ 
\eta \sum_j\left[ \theta _j\left( t\right) -\sum_kW_{0k}\left( t\right) \chi
_k\left( \alpha _j\right) \theta _k\left( t\right) \right] \theta _i\left(
t\right) \chi _i\left( \alpha _j\right) 
\end{array}
\end{equation}
$\gamma ,\eta >0$ . The intermediate layer nodes are equipped with a radial
basis function 
\begin{equation}
\label{2.2}\chi _i\left( \alpha \right) =\frac{e^{-\left( \alpha -\alpha
_i\right) ^2/2\sigma _i^2}}{\sum_{k=0}^Ne^{-\left( \alpha -\alpha _k\right)
^2/2\sigma _k^2}} 
\end{equation}
where in general we use $\sigma _i=\sigma $ for all $i$. The output is a
simple additive node.

The learning constant $\gamma $ should be sufficiently small to insure that
the learning time is much smaller than the characteristic times of the data $%
x(t)$. If this condition is satisfied each node parameter $\theta _i$ tends
to $\left\langle \cos \alpha _ix\right\rangle $, the real part of the
characteristic function $f(\alpha )$ for $\alpha =\alpha _i$.

The $W_{0i}$ learning law was chosen to minimise the error function 
\begin{equation}
\label{2.3}f\left( W\right) =\frac 12\sum_j\left( \theta _j-\sum_kW_{0k}\chi
_k\left( \alpha _j\right) \theta _k\right) ^2 
\end{equation}
One sees that the learning scheme is an hybrid one, in the sense that the
node parameter $\theta _i$ learns, in an unsupervised way, (the real part
of) the characteristic function $f(\alpha _i)$ and then, by a supervised
learning scheme, the $W_{0i}$'s are adjusted to reproduce the $\theta _i$
value in the output whenever the input is $\alpha _i$. Through the learning
law (\ref{2.1}) each node parameter $\theta _i$ converges to $\left\langle
\cos \alpha _ix\right\rangle $ and the interpolating nature of the radial
basis functions guarantees that, after training, the network will
approximate the real part of the characteristic function for any $\alpha $
in the domain $\left[ \alpha _0,\alpha _N\right] $.

A similar network is constructed for the imaginary part of the
characteristic function, where now 
\begin{equation}
\label{2.4}\theta _i\left( t+1\right) =\theta _i\left( t\right) +\gamma
\left( \sin \alpha _ix\left( t\right) -\theta _i\left( t\right) \right) 
\end{equation}

For higher dimensional data the scheme is similar. The number of required
nodes is $N^d$ for a $d$-dimensional data vector $x\left( t\right) $. For
example for the 2-dimensional data of Fig.1 we have used a set of $N^2$
nodes (Fig.4)

Each node in the square lattice has two inputs for the two components $%
\alpha _1$ and $\alpha _2$ of the vector argument of $f\left( 
\overrightarrow{\alpha }\right) $. The learning laws are, as before

\begin{equation}
\label{2.5}
\begin{array}{c}
\theta _{(ij)}\left( t+1\right) =\theta _{(ij)}\left( t\right) +\gamma
\left( \cos 
\overrightarrow{\alpha _{(ij)}}\overrightarrow{x\left( t\right) }-\theta
_{(ij)}\left( t\right) \right) \\ W_{0(ij)}\left( t+1\right)
=W_{0(ij)}\left( t\right) + \\ 
\eta \sum_{(kl)}\left[ \theta _{(kl)}\left( t\right)
-\sum_{(mn)}W_{0(mn)}\left( t\right) \chi _{(mn)}\left( \overrightarrow{%
\alpha _{(kl)}}\right) \theta _{(mn)}\left( t\right) \right] \theta
_{(ij)}\left( t\right) \chi _{(ij)}\left( \overrightarrow{\alpha _{(kl)}}%
\right) 
\end{array}
\end{equation}
The pair $(ij)$ denotes the position of the node in the square lattice and
the radial basis function is 
\begin{equation}
\label{2.6}\chi _{(ij)}\left( \alpha \right) =\frac{e^{-\left( 
\overrightarrow{\alpha }-\overrightarrow{\alpha _{(ij)}}\right) ^2/2\sigma
_{(ij)}^2}}{\sum_{(kl)}e^{-\left| \overrightarrow{\alpha }-\overrightarrow{%
\alpha _{(kl)}}\right| /2\sigma _{(kl)}^2}} 
\end{equation}
Two networks are used, one for the real part of the characteristic function,
another for the imaginary part with, in Eqs.(\ref{2.5}), $\cos 
\overrightarrow{\alpha _{(ij)}}\overrightarrow{x\left( t\right) }$ replaced
by $\sin \overrightarrow{\alpha _{(ij)}}\overrightarrow{x\left( t\right) }$.

Figs.5a-b shows the values computed by our algorithm for the real and
imaginary parts of the characteristic function corresponding to the
two-dimensional signal in Fig.1. On the left is a plot of the exact
characteristic function and on the right the values learned by the network.
In this case we show only the mesh corresponding to the $\theta _i$ values.
One obtains a 2.0\% accuracy for the real part and 4.5\% accuracy for the
imaginary part.

The convergence of the learning process is fast and the approximation is
reasonably good. Notice in particular the slope discontinuity at the origin
which reveals the non-existence of a second moment. The parameters used for
the learning laws in this example were $\gamma $=0.00036, $\eta $=1.8, $%
\sigma $=0.25. The number of points in the training set is 25000.

For a second example the data was generated by a Weierstrass random walk
with probability distribution 
\begin{equation}
\label{2.7}p(x)=\frac 16\sum_{j=0}^\infty \left( \frac 23\right) ^j\left(
\delta _{x,b^j}+\delta _{x,-b^j}\right) 
\end{equation}
and b=1.31, which is a process of the L\'evy flight type. The characteristic
function, obtained by the network, is shown in Fig. 6. Taking the $\log
\left( -\log \right) $of the network output one obtains the scaling exponent
1.49 near $\alpha $=0, close to the expected fractal dimension of the random
walk path (1.5). The parameters used for the learning laws in this example
were $\gamma $=0.0005, $\eta $=1.75, $\sigma $=0.1732. The number of points
in the training set is 80000.

These examples test the algorithm as a process identifier, in the sense
that, after the learning process, the network is a dynamical representation
of the characteristic function and may be used to perform all kinds of
analysis of the statistics of the data. There are other ways to obtain the
characteristic function of a probability distribution, which may be found in
the statistical inference literature (Prakasa Rao 1987). Our purpose in
developing neural-like algorithms for this purpose was both to have a device
that, after learning, is quick to evaluate and, at the same time, adjusts
itself easily, through continuous learning, to changing statistics. As the
PCA algorithms that extract the full correlation matrix, our neural
algorithm laws are also non-local. As a computation algorithm this is not a
serious issue, but for hardware implementations it might raise some problems.

\section{Identification of stochastic processes}

As we have stated before the full characterisation of the time structure of
a stochastic process requires the knowledge of its characteristic functional
(Eq. \ref{1.3}) for a dense set of functions $\xi (t)$.

To construct an approximation to the characteristic functional we might
discretize the time and the inner product in the exponential becomes a sum
over the process sampled at a sequence of times. 
\begin{equation}
\label{3.1}F\left( \xi \right) =\left\langle e^{i\sum_kx(t_k)\xi
(t_k)}\right\rangle 
\end{equation}
The problem would then be reduced to the construction of a multidimensional
characteristic function as in Section 2. In practice we would have to limit
the time-depth of the functional to a maximum of $T$ time steps, $T\Delta t$
being the maximum time-delay over which time correlations may be obtained.
If $N$ is the number of different $x$ values for each $k$, the algorithm
developed in Section 2 requires $N^T$ nodes in the intermediate layer and,
for any reasonably large $T$, this method becomes computationally explosive.

An alternative and computationally simpler method consist in, restricting
ourselves to Markov processes, to characterise the process by the
construction of networks to represent the transition function for fixed time
intervals. From these networks the differential Chapman-Kolmogorov equation
may then be reconstructed. Let $x(t)$ be a one dimension Markov process and $%
p\left( x_2,t+\Delta t|x_1,t\right) $ its transition function, that is, the
conditional probability of finding the value $x_2$ at time $t+\Delta t$
given $x_1$ at time $t$. Assume further that the process is stationary 
\begin{equation}
\label{3.2}p\left( x_2,t+\Delta t|x_1,t\right) =p\left( x_2,\Delta
t|x_1,t\right) 
\end{equation}

The network that configures itself to represent this function is similar to
the one we used for the 2-dimensional characteristic function. It is
sketched in Fig.s 7a-b. It has a set of $N^2$ intermediate layer nodes with
node parameters, the node with coordinates $\overrightarrow{x}_{(ij)}$
corresponding to the arguments $\left( x_2(ij)=x_0+i\Delta
x,x_1(ij)=x_0+j\Delta x\right) $ in the transition function. The domain of
both arguments is $\left( x_0,x_0+N\Delta x\right) $. For each pair $\left(
x_2=x(t+\Delta t),x_1=x(t)\right) $ in the data set, the node parameters
that are updated are those in the 4 columns containing the nearest
neighbours of the point $\overrightarrow{x}=\left( x_2,x_1\right) $ (Fig.
7b). The learning rule is 
\begin{equation}
\label{3.3}\theta _{(ij)}(t+1)=\frac{S_j(t)\theta _{(ij)}(t)+\frac{\eta
_{(ij)}(\overrightarrow{x})\exp \left( -\left| \overrightarrow{x}-%
\overrightarrow{x}_{(ij)}\right| ^2/\alpha \right) }{\sum_{(kl)}\eta _{(kl)}(%
\overrightarrow{x})\exp \left( -\left| \overrightarrow{x}-\overrightarrow{x}%
_{(kl)}\right| ^2/\alpha \right) }}{S_j(t+1)} 
\end{equation}
\begin{equation}
\label{3.4}S_j(t+1)=S_j(t)+\sum_i\frac{\eta _{(ij)}(\overrightarrow{x})\exp
\left( -\left| \overrightarrow{x}-\overrightarrow{x}_{(ij)}\right| ^2/\alpha
\right) }{\sum_{(kl)}\eta _{(kl)}(\overrightarrow{x})\exp \left( -\left| 
\overrightarrow{x}-\overrightarrow{x}_{(kl)}\right| ^2/\alpha \right) } 
\end{equation}
where $\eta _{(ij)}(\overrightarrow{x})=1$ if $(ij)$ is one of the nearest
neighbours of the data point and zero otherwise. $\alpha $ is a
neighbourhood smoothing factor. $S_j(t)$ is a column normalisation factor.
In the limit of large learning times the node parameters approach the
transition function 
\begin{equation}
\label{3.5}\theta _{(ij)}\rightarrow p\left( x_0+i\Delta x,\Delta
t|x_0+j\Delta x\right) 
\end{equation}
As for the networks in Section 2, the intermediate layer nodes are equipped
with a radial basis function (Eq. \ref{2.6}) and the connection strengths in
the output additive node have a learning law identical to the second
equation in (\ref{2.5}). The role of this part of the network is, as before,
to obtain an interpolating effect.

What the algorithm of Eqs.(\ref{3.3}) and (\ref{3.4}) does is to compute
recursively the average number of transitions between points in the
configuration space of the process. The spatial smoothing effect of the
algorithm automatically insures a good representation of a continuous
function from a finite data set. Furthermore its recursive nature would be
appropriate for the case of drifting statistics.

For a stationary process, once the learning process has been achieved and if 
$\Delta t$ is chosen to be sufficiently small, the network itself may be
used to simulate the stationary Markov process. A complete characterisation
of the process may also be obtained by training a few similar networks for
different (small) $\Delta t$ values and computing the coefficient functions
in the differential Chapman-Kolmogorov equation (Gardiner 1983). 
\begin{equation}
\label{3.6}
\begin{array}{c}
\partial _tp\left( 
\overrightarrow{z},t^{^{\prime }}|\overrightarrow{y},t\right) =-\sum_i\frac
\partial {\partial z_i}\left( A_i(\overrightarrow{z},t^{^{\prime }})p(%
\overrightarrow{z},t^{^{\prime }}|\overrightarrow{y},t)\right) \\ +\sum_{ij} 
\frac{\partial ^2}{\partial z_i\partial z_j}\left( B_{ij}(\overrightarrow{z}%
,t^{^{\prime }})p(\overrightarrow{z},t^{^{\prime }}|\overrightarrow{y}%
,t)\right) \\ +\int d\overrightarrow{x}\left\{ W(\overrightarrow{z}|%
\overrightarrow{x},t^{^{\prime }})p(\overrightarrow{x},t^{^{\prime }}|%
\overrightarrow{y},t)-W(\overrightarrow{x}|\overrightarrow{z},t^{^{\prime
}})p(\overrightarrow{z},t^{^{\prime }}|\overrightarrow{y},t)\right\} 
\end{array}
\end{equation}
The coefficients are obtained from the transition probabilities, noticing
that for all $\epsilon >0$%
\begin{equation}
\label{3.7}\lim _{\Delta t\rightarrow 0}p(\overrightarrow{x},t+\Delta t|%
\overrightarrow{z},t)=W(\overrightarrow{x}|\overrightarrow{z},t)\bigskip\ 
\bigskip\ for\bigskip\ \bigskip\ \left| \overrightarrow{x}-\overrightarrow{z}%
\right| \geq \epsilon 
\end{equation}
\begin{equation}
\label{3.8}\lim _{\Delta t\rightarrow 0}\frac 1{\Delta t}\int_{\left| 
\overrightarrow{x}-\overrightarrow{z}\right| <\epsilon }dx(x_i-z_i)p(%
\overrightarrow{x},t+\Delta t|\overrightarrow{z},t)=A_i(\overrightarrow{z}%
,t)+O(\epsilon ) 
\end{equation}
\begin{equation}
\label{3.9}\lim _{\Delta t\rightarrow 0}\frac 1{\Delta t}\int_{\left| 
\overrightarrow{x}-\overrightarrow{z}\right| <\epsilon
}dx(x_i-z_i)(x_j-z_j)p(\overrightarrow{x},t+\Delta t|\overrightarrow{z}%
,t)=B_{ij}(\overrightarrow{z},t)+O(\epsilon ) 
\end{equation}
$W(\overrightarrow{x}|\overrightarrow{z},t)$ is the jumping kernel, $A_i(%
\overrightarrow{z},t)$ the drift and $B_{ij}(\overrightarrow{z},t)$ the
diffusion coefficient.

As an example we have considered a Markov process with jumping, drift and
diffusion. A typical sample path is shown in Fig. 8. Three networks were
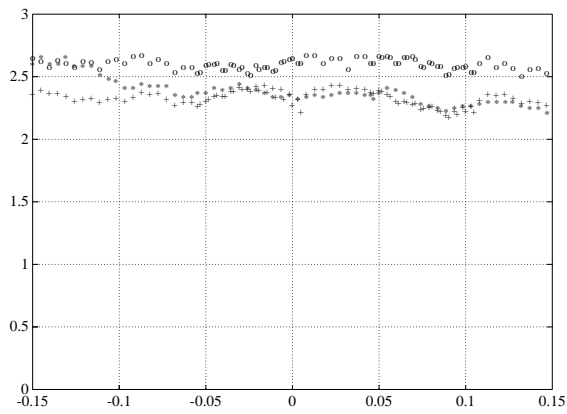
trained on this process, to learn the transition function for $t=\Delta t$, $%
2\Delta t$ and $3\Delta t$ ($\Delta t=0.374$ms). Fig. 9 shows the transition
function for $t=\Delta t$ and $3\Delta t$. Fig. 10 shows two sections of the
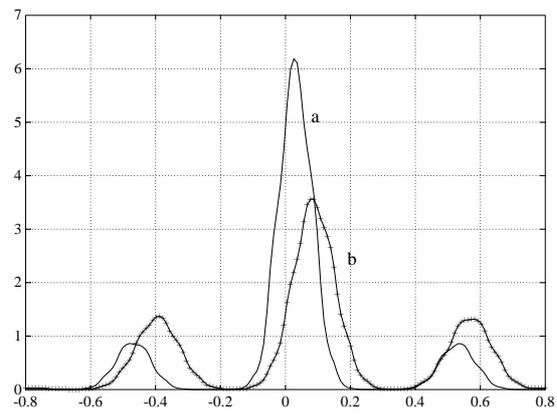
transition function for $x_2=0$, that is, $p(x,\Delta t|0,0)$ and $%
p(x,3\Delta t|0,0)$.

The networks were then used to extract the coefficient functions $A(x,t)$, $%
B(x,t)$ and $W(x|z,t)$. To find the drift $A(x,t)$ we use Eq. (\ref{3.8}).
Fig. 11 shows the computed drift function and a least square linear fit.
Also shown is a comparison with the exact drift function of the process.

To obtain the diffusion coefficient $B(x,t)$ we use (\ref{3.9}). Fig. 12
shows the diffusion coefficient for different $\Delta t$ values. $\Delta t$
is the smallest time step used in the process simulation. Therefore $%
B(x,t)=2.6$ is our estimate for the diffusion coefficient. In this case,
because the diffusion coefficient is found to be a constant, the value of
the jumping kernel $W(x|z,t)$ is easily obtained by integration around the
local maxima $x_m$ of $p(x,\Delta t|z)$ with $\left| x-z\right| >0.2$. 
\begin{equation}
\label{3.10}W=\frac 1{\Delta t}\int_{x_m-\delta }^{x_m+\delta }p(x,\Delta
t|z)dx 
\end{equation}
with $\delta =0.2$. We conclude $W(x|z)\cong 300\delta (\left| x-z\right|
-0.5)$. The parameters used for the learning laws in this example were $\eta 
$=0.48, $\alpha $=0.00021. The number of points in the training set is
1000000.

{\bf REFERENCES}

Dente, J. A. and Vilela Mendes, R. (1996), Unsupervised learning in general
connectionist systems, {\it Network: Computation in Neural Systems} 7, 123

Doyne Farmer, J. (1990), A Rosetta stone for connectionism, {\it Physica}
D42, 153

Gardiner, C. W. (1983), {\it Handbook of stochastic methods} (Springer,
Berlin)

Hebb, D. O. (1949), {\it The organisation of behaviour} (Wiley, New York)

Hida, T. (1980), {\it Brownian motion} (Springer, Berlin)

Lukacs, E. (1970), {\it Characteristic functions} (Griffin, London)

Oja, E. (1989), Neural networks, principal components and subspaces {\it %
Int. J. of Neural Systems} 1, 61

Oja, E. (1992), Principal components, minor components and linear neural
networks, {\it Neural Networks} 5, 927

Prakasa Rao, B. L. S. (1987), {\it Asymptotic theory of statistical inference%
} (Wiley, New York)

Sanger, T. D. (1989), Optimal unsupervised learning in a single-layer linear
feedforward neural network, {\it Neural Networks} 2, 459

Shlesinger, M. F., Zaslavski, G. M. and Klafter, J. (1993), Strange
kinetics, {\it Nature} 363, 31

Softky, W. R. and Kammen, D. M. (1991), Correlations in high dimensional or
asymmetric data sets: Hebbian neuronal processing, {\it Neural Networks} 4,
337

Taylor, J. G. and Coombes, S. (1993), Learning higher order correlations, 
{\it Neural Network}s 6, 423

Zumofen, G. and Klafter, J. (1993), Scale-invariant motion in intermittent
chaotic systems, {\it Physical Review} E47, 851

Zumofen, G. and Klafter, J. (1994), Random walks in the standard map, {\it %
Europhysics Letters} 25, 565

\end{document}